\documentclass[
 reprint,
 amsmath,amssymb,
 aps,floatfix,prl,superscriptaddress
]{revtex4-2}

\usepackage{braket}
\usepackage[normalem]{ulem} 

\usepackage{bbold}
\usepackage{graphicx}
\usepackage{enumerate}

\usepackage{tikz}
\usetikzlibrary{calc,positioning}

\newcounter{paranum}

\newcommand{\pnum}[1]{%
  \textit{#1}---}

\RequirePackage{color}
\definecolor{dkgreen}{rgb}{0.2,0.7,0.4}
\definecolor{dkblue}{rgb}{0.2,0.2,0.7}
\definecolor{dkred}{rgb}{0.8,0,0}
\definecolor{dkpurple}{rgb}{0.45,0.2,0.55}

\begin{document}
\title{Hilbert Space Fragmentation from Generalized Symmetries}
\author{Thea Budde}
\author{Marina Krist\'{c} Marinkovi\'{c}}
\author{Joao C. Pinto Barros}
\affiliation{Institut f\"{u}r Theoretische Physik, ETH Z\"{u}rich,
Wolfgang-Pauli-Str. 27, 8093 Z\"{u}rich, Switzerland}

\begin{abstract}
    Hilbert space fragmentation refers to exponential growth in the number of dynamically disconnected Krylov sectors with system size. It is taken as evidence of ergodicity breaking, since conventional symmetries generate at most a polynomial number of sectors. However, we demonstrate that generalized symmetries can fragment the Hilbert space. Models with higher-form, subsystem, and gauge symmetries can have exponentially many symmetry sectors. We further prove that non-invertible symmetries can induce additional fragmentation within individual symmetry sectors. Fragmentation in several known models arises from generalized symmetries, and the presence of exponentially many Krylov sectors therefore does not by itself imply ergodicity breaking. Finally, we show that disorder free localization arises naturally from Krylov-restricted thermalization when sectors lack translation invariance, requiring neither ergodicity breaking nor gauge symmetry.
\end{abstract}

\maketitle

\pnum{Introduction}
Isolated non-integrable quantum many-body systems are generically expected to be ergodic, such that local observables thermalize to a statistical ensemble consistent with the symmetries of the underlying system \cite{Deutsch:2018ulr,DAlessio:2015qtq,Gogolin:2015gts,Garrison:2015lva,Brighi:2022kca}, in line with the Eigenstate Thermalization Hypothesis (ETH)~\cite{PhysRevLett.54.1879,Srednicki:1994mfb}.
However, notable exceptions in models with constrained dynamics have been identified, including models with quantum many-body scars \cite{Pakrouski:2020hym,Banerjee:2020tgz,biswasScarsProtectedZero2022,Desaules:2022kse,wangInterrelatedThermalizationQuantum2023,Budde:2024rql,sauSublatticeScarsTwodimensional2024,Halimeh:2025vvp,hartseStabilizerScars2025,calajoQuantumManybodyScarring2025,miaoExactQuantumManyBody2025,guptaExactStabilizerScars2026,Budde:2026udh,Pal:2024omo} and systems with Hilbert space fragmentation \cite{moudgalya_quantum_2021, Sala_2020, Moudgalya_2022,Yang:2019mft,Rakovszky:2020apb,Steinegger2025,Mukherjee:2021iki,Khudorozhkov:2021oig,chattopadhyay2023strong,Tomasi:2019zkn,Kwan2025,Zhou:2026qwx,Ma:2026qnl,Han:2026rxg}.
These departures from ergodicity are not accounted for by conventional symmetries that are global and have a group structure \cite{moudgalya_quantum_2021}. 

Generalized symmetries \cite{Schafer-Nameki:2023jdn,Bhardwaj:2023kri,Kong:2020cie,Brennan:2023mmt,Choi:2022jqy, McGreevy:2022oyu} 
go beyond this framework and are characterized by topological operators that can be supported on submanifolds and may lack a group structure. They include higher-form symmetries \cite{Gaiotto:2014kfa,Cao:2023doz,Kapustin:2013uxa,Barkeshli:2023bta,Barkeshli:2022edm} 
and non-invertible symmetries \cite{Ortiz:2025psr, Giridhar:2025bol, Shao:2023gho,Zhu:2025ktk,Kaidi:2022cpf,Bhardwaj:2022lsg,Bhardwaj:2023ayw,Inamura:2023qzl,Inamura:2021szw,Moradi:2023dan,Seiberg:2024gek,Choi:2024rjm,Inamura:2026hif,Lootens:2021tet}. While generalized symmetries have proven powerful in constraining low-energy properties \cite{Thorngren:2019iar,Yoshida:2015cia,Cordova:2022ruw,Cordova:2017vab,Gaiotto:2017yup}, their implications for real-time dynamics \cite{Fukushima:2023svf,Stahl:2025zms,Stahl:2023tkh,sohal2025,Li:2025bgo,Steinegger2025,Fontana2026,Hart2022,Armas:2023tyx,Stephen2024} remain largely unexplored.

In this Letter, we demonstrate that generalized symmetries impact the real-time dynamics of quantum many-body systems 
and underlie effects previously interpreted as ergodicity breaking. We prove that both higher-form and non-invertible symmetries can generate exponentially many symmetry sectors in a tensor product basis, leading to Hilbert space fragmentation without necessarily implying ergodicity breaking. 
Individual sectors may nevertheless exhibit nonergodic dynamics, such as quantum many-body scarring.
This provides a unified framework of ergodicity-breaking phenomena across fragmented models, systems with generalized symmetries, and gauge theories. As a concrete example, we explain disorder free localization \cite{Smith_2017, Gyawali:2024hrz, Halimeh:2025vvp,Karpov:2020nhy,Halimeh:2021nep,Li:2021lay,Papaefstathiou:2020rvv} through Krylov-restricted thermalization. This work identifies generalized symmetries as the origin of a broad range of ergodicity-breaking phenomena. 

\pnum{Fragmentation from Symmetries}In fragmented models, dynamics initialized in a product state explore only a small part of the Hilbert space.
The dynamically accessible subspace from an initial state $\ket{\psi_0}$ under unitary evolution is its Krylov sector,
\begin{equation}
    \mathcal{K}(\ket{\psi_0})
    \equiv \mathrm{span} \left\{ \ket{\psi_0}, H \ket{\psi_0}, H^2 \ket{\psi_0}, \ldots \right\}.
\end{equation}
Fragmentation refers to exponential growth with system size in the number of distinct Krylov sectors generated from a tensor product basis \cite{moudgalya_quantum_2021,Sala_2020,Moudgalya_2022}. Any symmetry diagonal in the tensor product basis constrains $\mathcal{K}(\ket{\psi_0})$ to a symmetry sector. Because conventional symmetries generate at most polynomially many such sectors \cite{moudgalya_quantum_2021}, fragmentation has been taken as evidence of ergodicity breaking.

Recent work showed that $(2+1)$-dimensional models with $U(1)$ 1-form symmetries exhibit an exponentially growing number of Krylov sectors \cite{sohal2025}. We generalize this result by showing that a broad class of conserved operators supported on subregions of the lattice leads to fragmentation. The following proposition establishes sufficient conditions under which such conserved operators generate exponentially many Krylov sectors.

Proposition I: Consider a local translation-invariant Hamiltonian $H$ on a $d$-dimensional hypercubic lattice of linear size $L$ with periodic boundary conditions. The Hilbert space is a tensor product 
$\mathcal{H}= \bigotimes_i h$ of identical local Hilbert spaces $h$.
Assume there exists an operator $U$ that
\begin{enumerate}[i)]
    \item is conserved, so $[H, U] = 0$,
    \item admits $t(L) \geq a L^f$ translations $U_\alpha = T^{-1}_\alpha U T_\alpha$ that have disjoint support, for a specific $f \geq 1$, $a>0$ and all sufficiently large $L$, and
    \item is diagonal in a product basis $U \ket{\phi_j} = g_{j} \ket{\phi_j}$ with a finite number of distinct eigenvalues $q \geq 2$,
\end{enumerate}
then the number of Krylov sectors in the basis $\ket{\phi_j}$ is at least $\exp(cL^f)$ for some $c > 0$ and all sufficiently large $L$. Thus, the model exhibits Hilbert space fragmentation of the tensor product basis. 

Proof: Translation invariance of the Hamiltonian implies that $[U_\alpha, H] = 0$ for all $\alpha$. By assumption (ii), we can choose $t(L)$ of these translated operators with pairwise disjoint support. By (iii), all $U_\alpha$ are diagonal in the same product basis and admit $q$ eigenvalues on product states. Since the Hilbert space factorizes, product states can realize arbitrary combinations of these eigenvalues. Using $t(L)\ge aL^f$, this yields at least $q^{t(L)}\ge \exp(cL^f)$ symmetry sectors with $c=a\log q>0$. Finally, because the translated operators $U_\alpha$ are diagonal in the chosen product basis, each Krylov sector is contained within an individual symmetry sector. Hence, the total number of Krylov sectors is at least $\exp(cL^f)$. This proves Hilbert space fragmentation of Hamiltonians that fulfill the conditions of Proposition I.

Although open boundary conditions break translation symmetry, if the Hamiltonian consists of translations of a fixed local term plus boundary contributions, the counting argument is unchanged: one can place $\mathcal{O}(t(L))$ spatially shifted copies of $U$ with disjoint support in the bulk, yielding the same scaling up to subextensive boundary corrections.

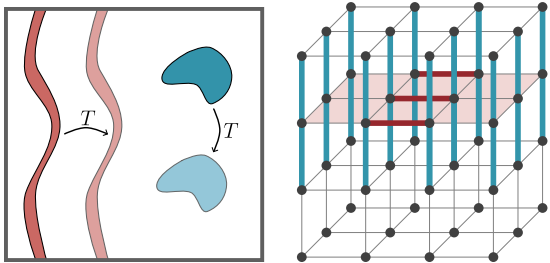
\begin{figure}
    \centering
        \includegraphics[width=0.4\linewidth]{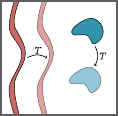}
        \hspace{0.02\linewidth}
    \resizebox{!}{0.4\linewidth}{
            
            \begin{tikzpicture}
            \definecolor{ethblue}{RGB}{51, 149,171}

            \definecolor{ethred}{RGB}{150, 39, 45}

            \definecolor{ethgreen}{RGB}{241, 215, 213}

            \def \dx{2};
            \def \dy{2.1};
            \def \dz{2};
            \def \nbx{4};
            \def \nby{4};
            \def \nbz{3};

            \fill[ethgreen]
                (\dx,      3*\dy, \dz)
             -- (\nbx*\dx, 3*\dy, \dz)
             -- (\nbx*\dx, 3*\dy, \nbz*\dz)
             -- (\dx,      3*\dy, \nbz*\dz)
             -- cycle;
            
            \foreach \x in {1,...,\nbx} {
                \foreach \z in {1,...,\nbz}{
                    \draw [gray] (\x*\dx,\dy,\z*\dz) -- ( \x*\dx,\nby*\dy,\z*\dz);
                }
            }
            
            \foreach \y in {1,...,\nbx} {
                \foreach \z in {1,...,\nbz}{
                    \draw  [gray] (\dx,\y*\dy,\z*\dz) -- ( \nbx*\dx,\y*\dy,\z*\dz);
                }
            }
            
            \foreach \x in {1,...,\nbx} {
                \foreach \y in {1,...,\nby}{
                    \draw  [gray] (\x*\dx,\y*\dy,\dz) -- ( \x*\dx,\y*\dy,\nbz*\dz);
                }
            }

            \foreach \x in {1,...,\nbx} {
            \foreach \z in {1,...,\nbz} {
                \draw[ethblue, line width=5] (\x*\dx,2*\dy,\z*\dz) -- ( \x*\dx,4*\dy,\z*\dz);
            }}

            \foreach \z in {1,...,\nbz} {
                \draw[ethred, line width=5] (2*\dx,3*\dy,\z*\dz) -- ( 3*\dx,3*\dy,\z*\dz);
            }

            
            \foreach \z in {1,...,\nbz} {
                \draw[ethblue, line width=5] (3*\dx,2*\dy,\z*\dz) -- ( 3*\dx,4*\dy,\z*\dz);
            }
            
            \foreach \x in {1,...,\nbx} {
                \foreach \y in {1,...,\nby} {
                    \foreach \z in {1,...,\nbz} {
                        \node [gray] at (\x*\dx,\y*\dy,\z*\dz) [circle, fill=darkgray] {};
                    }
                }
            }
            \end{tikzpicture}
        }
    
    \caption{Left: The group elements of higher-form symmetries (red), and gauge symmetries (blue) have support on a subregion of the lattice. Their disjoint translations are also independent symmetries of the translation-invariant Hamiltonian. This leads to fragmentation of the tensor product Hilbert space. Right: The 2-form non-invertible symmetry $D_{xy}$ in Eq.~\eqref{eq:noninvertible} projects to sectors in which the spins in two neighboring planes (blue) are all equal. The dynamics in the highlighted plane are then decoupled. Here, a winding symmetry measuring the magnetization of the spins in one row or column (red) is conserved. Therefore, $D_{xy}$ commutes with the Hamiltonian.}
    \label{fig:fragFromSym}
\end{figure}

Not all conserved operators that fulfill these conditions are typically classed as symmetries. Additional requirements, such as locality or topology, are typically required for an operator to be a symmetry, depending on the context. But there are multiple well-known classes of symmetries whose symmetry operators exhibit these properties. We illustrate their geometries in Fig.~\ref{fig:fragFromSym}. 

The first class of symmetries that fulfill the properties of Proposition I is gauge symmetries. Their symmetry operators are $k$-local and therefore have on the order of $L^d$ disjoint translations. The number of superselection and Krylov sectors then scales with at least $\exp(c L^d)$. 
Gauge theories are defined as constrained systems on a single sector of this symmetry. However, taking the full structure of the product basis into account is important when performing quantum simulations of these theories in which noise may weakly couple to the remaining sectors \cite{Halimeh:2025vvp}. It is widely established that the number of symmetries in gauge theories with finite-dimensional local Hilbert spaces grows with system size, making it possible for them to exhibit exponentially many superselection sectors \cite{Ciavarella:2025zqf,Steinegger2025}.
The fact that the dimension of the largest superselection sector typically grows exponentially slower than that of the product Hilbert space \cite{Desaules:2022kse,Luo:2020stn,Mariani:2023eix} implies that fragmentation from gauge symmetry is typically strong.

The second class of symmetries that fulfill Proposition I is higher-form symmetries \cite{Gaiotto:2014kfa,Cao:2023doz}. The generator of a $p$-form symmetry has support on a $(d-p)$-dimensional submanifold of the lattice, where $d$ is the spatial dimension of the system. Higher-form symmetries are topological, meaning they become invariant under deformations in at least one superselection sector. Higher-form symmetries that are not topological are called subsystem higher-form symmetries \cite{Cao:2023doz}. There are on the order of $L^p$ disjoint translations of a $p$-form symmetry, since the operator can be translated in $p$ directions. Thus, if the symmetry is diagonal in the product basis, there are at least $\exp(c L^p)$ Krylov sectors of the tensor product basis. 
Higher-form symmetry, therefore, leads to fragmentation even beyond $U(1)$ $p$-form symmetries, which had been argued in \cite{sohal2025}. The symmetries observed in the fragmented system studied in \cite{Khudorozhkov:2021oig} are also 1-form. While higher-form symmetries fragment the product Hilbert space, they do not necessarily fragment individual superselection sectors of a gauge symmetry.

Other subsystem symmetries, whose support does not fit the standard $p$-form geometry \cite{Devakul_2019}, may still allow extensive disjoint translations and lead to fragmentation.

Not all gauge, higher-form, or subsystem symmetries need to have an operator that is diagonal in a product basis. However, for local symmetries, this is common.  For example, all quantum link models, including those with non-abelian symmetries \cite{Wiese:2021djl}, satisfy Proposition I for at least one operator.

\pnum{Fragmentation from non-invertible symmetry}Even when a symmetry fragments the product basis, individual sectors of other symmetries do not necessarily fragment. However, if a symmetry sector allows for a tensor product basis, we can predict its fragmentation under the following conditions.

Proposition II: Take a Hamiltonian $H$, a Hilbert space $\mathcal{H}$ and an operator $U$ with all conditions, except i), of Proposition I fulfilled.
Now define a projector $P$ onto a fixed symmetry sector $P\mathcal{H}$. If
\begin{enumerate}[i)]
    \setcounter{enumi}{3}
    \item $P$ and $UP$ are diagonal in a tensor product basis of $\mathcal{H}$, $P\mathcal{H}$ factorizes as a tensor product of local Hilbert spaces, and $[H, UP]=0$,
    \item the symmetry sector $P\mathcal{H}$ is translation invariant under the translations $T_\alpha$ and
    \item $U$ takes a finite number $q\geq 2$ of distinct eigenvalues in $P\mathcal{H}$,
\end{enumerate}
then an analogous counting argument of symmetry sectors goes through. The symmetry sector $P\mathcal{H}$ is therefore fragmented.
This generalizes the class of symmetries that can lead to fragmentation. It is not necessary for $U$ to commute with $H$, as long as $[H,UP]=0$. Thus, such partial isometries can also lead to fragmentation. 

Generic partial isometries \cite{Ortiz:2025psr, Giridhar:2025bol} are operators of the form $D = UP$, where $U$ is unitary and  $P$ is a positive semidefinite operator that acts as the identity in $P\mathcal{H}$. $D$ is unitary only within that subspace
\begin{equation}
    D^\dagger D = P^2.
\end{equation}
If $P\mathcal{H}$ is composed of a set of Krylov sectors, operators that take the form of partial isometries are compatible with the preservation of probability amplitudes within a Krylov sector following a generalized Wigner's theorem \cite{Ortiz:2025psr}. 
To preserve a notion of locality for symmetries, we restrict $P$ to be a projector onto a subset of sectors of an established symmetry.

The non-invertible symmetries in lattice Hamiltonians with Kramers-Wannier dualities are described by partial isometries \cite{Ortiz:2025psr, Giridhar:2025bol, Shao:2023gho, Zhu:2025ktk, Seiberg:2023cdc,Seiberg:2024gek}.
For example, the non-invertible symmetry operator in the critical Ising model decomposes into a product of a projector $(1+\eta)/2$ onto the even sector of the parity symmetry $\eta$, and a unitary operator \cite{Seiberg:2024gek}.

\pnum{Fragmentation in a Quantum Link Model}Here, we illustrate the mechanisms of fragmentation from generalized symmetries discussed above with an example. We show that the fragmentation observed in Ref.~\cite{Steinegger2025} is explained by a 2-form partial isometry. 
We begin by summarizing the results of Ref.~\cite{Steinegger2025}, which studies the 3-dimensional $U(1)$ pure-gauge quantum link model (QLM) on the cubic lattice with $S=1/2$
\begin{equation}
    H = - J \sum_\Box  \left( U_\Box + U_\Box^\dagger \right) + \lambda \sum_\Box  \left( U_\Box + U_\Box^\dagger \right)^2.
\end{equation}
The plaquette operator $U_{\Box}$ is the oriented product of link raising/lowering operators around an elementary plaquette $\Box$ in the $\mu$--$\nu$ plane 
$U_\Box = S^+_{r,\hat{\mu}} S^+_{r + \hat{\mu},\hat{\nu}} S^-_{r + \hat{\nu},\hat{\mu}} S^-_{r,\hat{\nu}}$
where $\hat{\mu}$ denotes the unit vector along the lattice direction $\mu$.
The model has local $U(1)$ gauge symmetry generated by the operators
$G_r = \sum_{\hat{i} \in (\hat{x},\hat{y},\hat{z})} \left(S^z_{r,\hat{i}} -  S^z_{r-\hat{i},\hat{i}}\right)$.
Since this symmetry is 2-local, takes 7 possible values and allows for $t(L) = L^3/2$ disjoint translations, it therefore has at least $7^{\frac{L^3}{2}}$ symmetry sectors in the $S^z$ basis. The sector in which $G_r \ket{\psi_\mathrm{phys}} = 0$ for all sites $r$ is called the physical sector.
Additionally, there are subsystem $1$-form symmetries given by planar magnetization operators
\begin{equation}
    W^{(1)}_{xy}(z) = \sum_{x,y} S^z_{r,\hat{z}},
\end{equation}
with support on $x$--$y$ planes, as denoted by the subscript, at fixed coordinate $z$. Here, we chose one set of directions $(x,y,z)$ for clarity, but all of the following analysis also holds for permutations of these directions. Furthermore, we included the upperscript $^{(1)}$ to emphasize that this is a 1-form symmetry. $W^{(1)}_{xy}(z)$ has $L^2+1$ distinct eigenvalues in the electric field basis and is local on a plane, allowing for $t(L) = L$ disjoint translations. By Proposition I, there are $(L^2+1)^L$ symmetry sectors in the full $S^z$ product basis, and the fragmentation is strong, since the dimension of its largest sector grows exponentially slower than that of the product Hilbert space. This symmetry is topological in the physical sector and cannot cause the individual sector to fragment.

Ref.~\cite{Steinegger2025} studies the physical sector restricted to the ``maximal-winding'' sector in which $W^{(1)}_{xy}(z)$ takes its maximal value on every $x$-$y$ plane. In this sector, the model reduces to $L$ copies of the $(2+1)$-dimensional QLM in the $x$-$y$ planes. Remarkably, this means that this symmetry sector itself fragments into exponentially many Krylov components, since the winding symmetries of the $(2+1)$-dimensional QLM
\begin{equation}
    W^{(2)}_{x}(y,z) = \sum_{x } S^z_{r,\hat{y}}
\end{equation}
are conserved in every plane independently.
This operator does not commute with $H$ on the full Hilbert space of the $(3+1)$-dimensional model. It is therefore not a symmetry, and has been interpreted as ergodicity breaking. 

We demonstrate that the additional Krylov sectors arise from a non-invertible symmetry. 
In the full product basis, where we do not restrict to the physical sector, the operator $ W^{(2)}_{x}(y,z)$ is conserved in many more sectors, since it suffices that the two adjacent planes have maximal or minimal winding. We therefore define an operator that projects $W^{(2)}_{x}(y,z)$ onto only the sectors in which it is conserved
\begin{equation}\label{eq:noninvertible}
D_{xy}(y,z, \alpha) \equiv e^{i \alpha W^{(2)}_{x}(y,z)} \, P_{xy}^{(\mathrm{max})}(z),
\end{equation} 
where $P_{xy}^{(\mathrm{max})}(z)$ is the projector onto the $W^{(1)}_{xy}(z) \ket{\psi} = \pm W^{(1)}_{xy}(z - 1) \ket{\psi} = \pm \frac{L^2}{2} \ket{\psi}$ sector. The geometry of $D_{xy}(y,z, \alpha)$ is illustrated in Fig.~\ref{fig:fragFromSym}.

The operator $D_{xy}(y,z, \alpha)$ fulfills most of the properties of a $U(1)$ symmetry, as
\begin{align}
    [H,D_{xy}(y,z, \alpha)] &=0, \\ 
    D_{xy}(y,z, \alpha) D_{xy}(y,z, \alpha') &= D_{xy}(y,z, \alpha + \alpha').
\end{align}
However, we can show that it is non-invertible on the full Hilbert space
\begin{equation}
    D_{xy} D_{xy}^\dagger = P_{xy}^{(\mathrm{max})}(z),
\end{equation}
where we use a shorthand notation $D_{xy} \equiv D_{xy}(y,z, \alpha)$. $D_{xy}$ is therefore a partial isometry. 

$P\mathcal{H}$ is a product Hilbert space, with some frozen factors, and $D_{xy}$ fulfills the conditions of Proposition II. Its existence, therefore, predicts the exponentially growing number of Krylov sectors that can be labeled by the eigenvalues of $D_{xy}$ in each $xy$-slice.

\pnum{Local symmetry of the PXP model}Originally studied to describe the dynamics of the Rydberg blockade, the PXP model \cite{Serbyn:2020wys, Bluvstein:2020ddz} is a paradigmatic model for ergodicity breaking in the context of quantum many-body scars \cite{Bernien:2017ubn, Turner:2017fxc, moudgalya_quantum_2021}. Additionally, it is important in the field of quantum simulations of gauge theories, since, when properly constrained, it maps to a $U(1)$ gauge theory in the physical sector \cite{Surace_2020}. Here, we show that the fragmented product Hilbert space can be predicted through Proposition I.
The PXP model is given by a $S=1/2$ spin chain with  Hamiltonian
\begin{equation}
    H = \sum_{n=1}^L P_n \sigma^x_{n+1} P_{n+2}
    \label{eq:PXP}
\end{equation}
where $P_n = \ket{\downarrow}\bra{\downarrow}_n$ projects site $n$ on the spin-down state. For simplicity, we adopt periodic boundary conditions, though the analysis can be straightforwardly extended to other choices. The model is typically restricted to the physical Hilbert space, which excludes states that have neighboring sites in the up-state $\ket{\uparrow\uparrow}_{n,n+1}$. This constraint does not change the physics, since the constrained Hilbert space is a Krylov sector. The constrained model describes the Rydberg blockade regime, in which $\ket{\downarrow}$ is interpreted as the ground state and $\ket{\uparrow}$ as the Rydberg state. 

When the same Hamiltonian is viewed on the full tensor product Hilbert space $\{\ket{\uparrow}, \ket{\downarrow}\}^{\otimes L}$, any pair of neighboring excitations $\ket{\uparrow \uparrow}$ is frozen under time evolution. Since each combination of adjacent excitations constitutes a distinct Krylov sector, the number of Krylov sectors grows exponentially. The PXP model is therefore a simple example of Hilbert space fragmentation \cite{moudgalya_quantum_2021}.

We connect this fragmentation with the presence of a local conserved operator.
The 2-local projector on neighboring excitations
\begin{equation}
    G_n = \ket{\uparrow\uparrow}\bra{\uparrow\uparrow}_{n,n+1}
\end{equation} 
commutes with $H$ and is conserved. 
This 2-local operator fulfills Proposition I and predicts fragmentation. Any other model with locally frozen states has an analogous local conserved quantity given by the projector on that state.

The existence of the conserved operator $G_n$ allows for the interpretation of the PXP model as an unconventional $\mathbb{Z}_2$ gauge theory. While there are no degrees of freedom that we can associate with gauge links or matter sites, we can define an operator $\tilde{G}_n = 1-2G_n$, which satisfies $\tilde{G}_n^2 = 1$. From this perspective, the constrained PXP model describing the Rydberg blockade regime corresponds to the Hamiltonian in Eq.~\eqref{eq:PXP} in the Gauss' law sector $\tilde{G}_n \ket{\psi_\mathrm{phys}} = \ket{\psi_\mathrm{phys}}$ for all $n$. It would be interesting to determine whether the PXP model can be obtained from a process that involves gauging a global $\mathbb{Z}_2$ symmetry. 


\pnum{Krylov-restricted thermalization}Since unitary dynamics does not mix distinct Krylov sectors, thermalization can only occur within each sector. This is known as Krylov-restricted thermalization \cite{Moudgalya:2019vlp,moudgalya_quantum_2021}. An analogous structure arises in models with generalized and gauge symmetries, where dynamics are confined to superselection sectors. In gauge theories, it is standard to resolve these symmetries by constraining the Hilbert space to a single physical sector. 
Ergodicity may nevertheless break within individual sectors. Integrable sectors \cite{HerzogArbeitman2019} and sectors exhibiting quantum many-body scars have been identified in many models with gauge symmetry \cite{Banerjee:2020tgz,biswasScarsProtectedZero2022,Desaules:2022kse,wangInterrelatedThermalizationQuantum2023,Budde:2024rql,sauSublatticeScarsTwodimensional2024,Halimeh:2025vvp,hartseStabilizerScars2025,calajoQuantumManybodyScarring2025,miaoExactQuantumManyBody2025,guptaExactStabilizerScars2026,Budde:2026udh}. Similar phenomena occur in fragmented models whose Krylov structure has not been attributed to symmetries \cite{moudgalya_quantum_2021, Ganguli2025}, suggesting that the structure of the fragmented Hilbert space provides a common mechanism for ergodicity breaking within sectors.

Identifying superselection sectors with Krylov sectors allows us to interpret disorder free localization \cite{Smith_2017, Gyawali:2024hrz, Halimeh:2025vvp,Karpov:2020nhy,Halimeh:2021nep,Li:2021lay,Papaefstathiou:2020rvv,Jeyaretnam_2025} as a consequence of Krylov-restricted thermalization. In sector-restricted dynamics, systems need not relax to translation-invariant ensembles, even in the absence of ergodicity breaking. For example, in gauge theories, an initial state with spatially inhomogeneous eigenvalues of the symmetry operators must retain this inhomogeneity under time evolution and therefore cannot relax to a translation-invariant ensemble (see Supplemental Material~\cite{suppmat}).
Disorder free localization in gauge theories arises when the initial state is a superposition over many Gauss' law sectors with spatially varying expectation values $\langle G_n \rangle$. Long-time dynamics then produce an average over inhomogeneous sector ensembles, which can appear to have a localized defect. Furthermore, if many sectors contain frozen degrees of freedom that suppress transport, sector-averaged correlation functions remain strongly suppressed at long times, as observed in Refs.~\cite{Karpov:2020nhy,Smith:2018smg}.
The same mechanism applies in fragmented models without gauge symmetry when Krylov-restricted ensembles are not translation-invariant. In this context, the resulting phenomenology is referred to as quasi-localization \cite{Sala_2020, Moudgalya_2022}.

\pnum{Conclusion}We have shown that generalized symmetries play a central role in the non-equilibrium dynamics of quantum many-body systems. In particular, broad classes of symmetries, including subsystem, higher-form, and gauge symmetries, as well as their non-invertible counterparts, generate exponentially many Krylov sectors of the tensor product Hilbert space.
Under the standard definition \cite{sohal2025, moudgalya_quantum_2021}, this constitutes Hilbert space fragmentation and therefore ergodicity breaking. However, in many cases, including a fragmented three-dimensional quantum link model \cite{Steinegger2025} and the PXP model \cite{moudgalya_quantum_2021}, these Krylov sectors coincide with symmetry sectors of generalized symmetries. If one widens the definition of symmetry to include these generalized symmetries, this no longer constitutes ergodicity breaking. We therefore suggest reserving the term Hilbert space fragmentation for models in which the Krylov structure cannot be explained by generalized symmetries.

We have demonstrated that resolving generalized symmetries explains the presence of an exponentially growing number of Krylov sectors in a broad class of systems relevant to condensed matter and high energy physics.  It would be interesting to understand whether all such instances can be explained by generalized symmetries. It has been proposed that the commutant algebra, the space of operators that commute with each local term in the Hamiltonian, is the required set of conserved quantities characterizing fragmented models \cite{Moudgalya_2022}. A key question is whether this algebra can always be spanned by generalized symmetries. If so, generalized symmetries would provide a unified explanation of fragmentation. If not, they can still provide further insight and a tractable approximation when the full commutant algebra cannot be determined.

We have also shown that disorder free localization \cite{Smith_2017, Gyawali:2024hrz, Halimeh:2025vvp,Karpov:2020nhy,Halimeh:2021nep,Li:2021lay,Papaefstathiou:2020rvv} does not require ergodicity breaking or gauge symmetry. Instead, it arises from Krylov-restricted dynamics, which can lead to non-translation-invariant thermal ensembles and suppressed transport when Krylov sectors themselves lack translation invariance.

Generalized symmetries provide a framework for identifying and understanding novel anomalous out-of-equilibrium dynamics in quantum many-body systems. Unlike systems with conventional symmetries, the number of conserved quantities in systems with higher form and gauge symmetries grows with system size, leading to many small symmetry sectors.
Non-invertible symmetries realized as partial isometries naturally describe quantities conserved only within subsets of symmetry sectors, thereby accounting for apparent strong ergodicity breaking. It is reasonable to expect that unresolved partial isometries may likewise give rise to apparent weak ergodicity breaking, such as quantum many-body scars, and thus offer new mechanisms for their emergence.

\section{Acknowledgements}

We thank Tin Sulejmanpasic for pointing out that the symmetry of the model discussed in Ref.~\cite{Khudorozhkov:2021oig} is higher form, inspiring the generalization presented here. We also thank Debasish Banerjee, Andrea Bulgarelli, Klemen Kersic, Kiryl Pakrouski, Zlatko
Papic, Yarden Scheffer, Arnab Sen, and Weronika Wiesiolek for enlightening
discussions. This research was supported in part by grant NSF PHY-2309135 to the Kavli Institute
for Theoretical Physics (KITP). TB thanks the organizers of the Computational Methods in Lattice Field Theories school at TIFR and the Exact Solvability and Quantum Information school at Les Houches for their hospitality and for providing a stimulating scientific environment where part of this work was developed. MKM is grateful for the hospitality of Perimeter Institute, where part
of this work was carried out. Research at Perimeter Institute is supported in part by the Government
of Canada through the Department of Innovation, Science and Economic Development and by
the Province of Ontario through the Ministry of Colleges and Universities. This research was also
supported in part by the Simons Foundation through the Simons Foundation Emmy Noether Fellows
Program at Perimeter Institute.

\bibliography{references}

\section{Supplemental Material}

\subsection{Examples of Krylov restricted Thermalization}


An individual symmetry or Krylov sector does not necessarily thermalize. But even if it does, its long-term dynamics may appear unusual. 
For thermal ensembles calculated over the full product basis, entropic order is prohibited \cite{Han:2025eiw}. This means that there is a temperature above which the ensemble cannot be ordered, and therefore has to be translation-invariant if the Hamiltonian is. 
This need not be the case in Krylov-restricted ensembles
\begin{equation}
    \langle\mathcal{O}\rangle = \frac{\mathrm{Tr}_\mathcal{K}\left(\mathcal{O}e^{-\beta H}\right)}{\mathrm{Tr}_\mathcal{K}\left(e^{-\beta H}\right)} = \frac{\sum_{\ket{\phi_j}\in\mathcal{K}} \bra{\phi_j}\mathcal{O}e^{-\beta H}\ket{\phi_j}}{\sum_{\ket{\phi_j}\in\mathcal{K}} \bra{\phi_j}e^{-\beta H}\ket{\phi_j}}.
\end{equation}
where the sum goes over the product basis states in the Krylov sector.

\begin{figure}[t]
    \centering
    \includegraphics[width=0.7\linewidth]{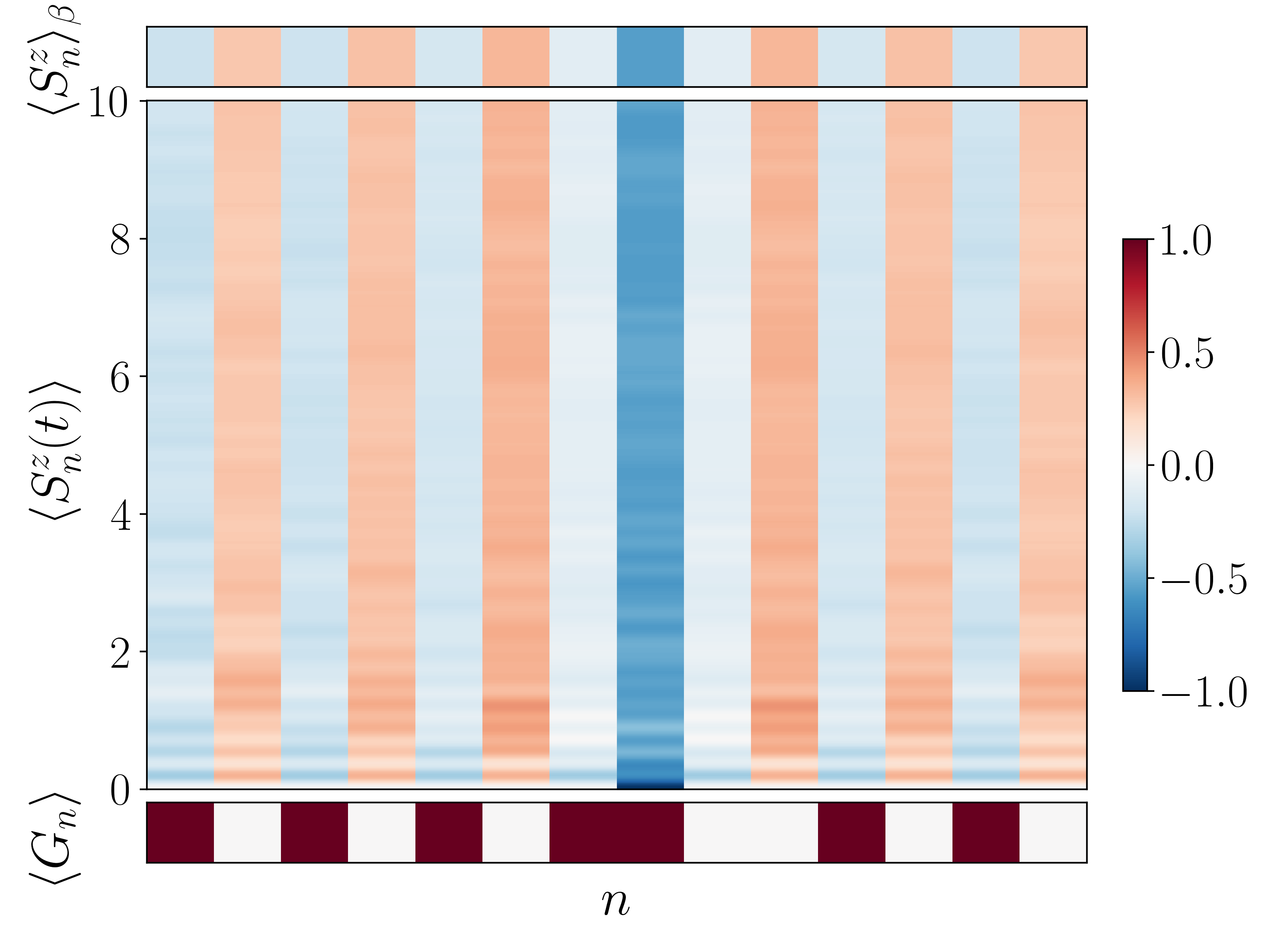}
    \caption{Fragmentation from local symmetry does not imply non-thermal dynamics, since thermalization can still occur within symmetry sectors. The figure illustrates the time evolution of the local magnetization of the quantum link model Eq.~\eqref{eq:1dqlm} with $L=14$ from an initial state with inhomogeneous Gauss' law detailed in Eq.~\eqref{eq:inogauss1}. Although local observables relax, the late-time profile remains non–translation-invariant, and matches the corresponding Krylov-restricted thermal prediction (top strip). The defect lies near the position of the defect in Gauss' law (bottom strip).
    }
    \label{fig:QLMThermalizing}
\end{figure}

We illustrate that sector-restricted ensembles may thermalize into non-translation-invariant ensembles using a gauge-theory example.
Fig.~\ref{fig:QLMThermalizing} illustrates the unitary dynamics, from a product state, of  the massless
$(1+1)$-dimensional, $S=1$, $U(1)$ quantum link model \cite{Halimeh:2025vvp}. Concretely, the Hamiltonian can be written as
\begin{equation} \label{eq:1dqlm}
    H=\sum_{n=1}^L \frac{\kappa}{2a}\left(c_n^\dagger S^+_n c_{n+1} + h.c.\right) + \frac{a}{2}\sum_{n=1}^L \left(S^z_n\right)^2.
\end{equation}
The fermionic operators $c_n$ live at the matter sites, while the spin operators $S_n^\alpha$ live at the links that connect the site $n$ to $n+1$. The parameters of the theory are $L$, which is the system size, $\kappa$ and $a$. We consider periodic boundary conditions, making this model translation invariant. Beyond translations, this model is also invariant under charge conjugation and reflection \cite{Banerjee2012}. Furthermore, it is a $U(1)$ gauge theory, reflected by the fact that the Hamiltonian commutes with a set of local generators defined by 
\begin{equation}
    G_n = c^\dagger_{n} c_{n} - S^z_{n} + S^z_{n-1}.
\end{equation}

To illustrate how this model can thermalize into a thermal state that breaks translation invariance without breaking ergodicity, we choose the parameters $L=14$, $\kappa = 1$ and $a = 0.1$. With exact diagonalization, we evolve from an initial state with an inhomogeneous Gauss' law pattern. We initialize the system in a product basis state $\ket{\psi}$ satisfying
\begin{equation}
    S^z_n\ket{\psi}=-\delta_{n8}\ket{\psi},\quad c^\dagger_nc_n\ket{\psi}=\left(\frac{1-(-1)^n}{2}\right)\ket{\psi}.
\end{equation}
This is a state that lies in the sector characterized by
\begin{equation}
    G_n\ket{\psi}=\left(\frac{1-(-1)^n}{2}+\delta_{n8}-\delta_{n9}\right)\ket{\psi}.
    \label{eq:inogauss1}
\end{equation}
There is an odd-even asymmetry that breaks translation invariance by one site, and Kronecker-Delta terms that make sites around $n=8$ special, breaking translation invariance entirely in this Krylov sector.
For this specific state, the Krylov restricted thermal ensemble is obtained by choosing $\beta = 0.00146$ such that $\langle H\rangle_\beta = a/2$ matches the initial energy of the initial state. The defect of the inhomogeneous Gauss' law remains visible at late times and is reproduced by the Krylov-restricted thermal ensemble even for $\beta \to 0$. 

\begin{figure}[t]
    \centering
    \includegraphics[width=0.7\linewidth]{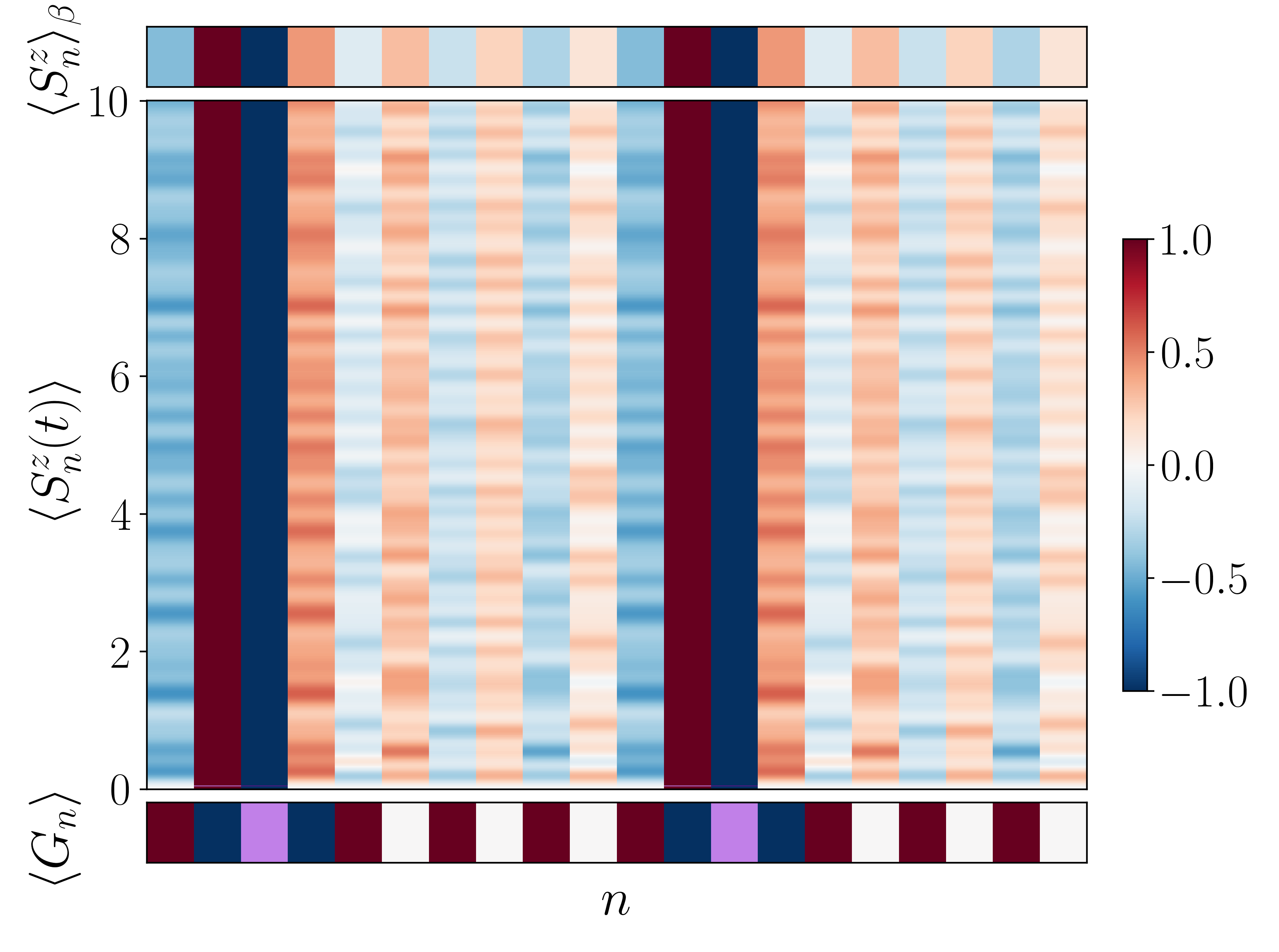}
    \caption{Local symmetries can lead to Krylov sectors that decouple the system into different components. The figure shows the time evolution of the local magnetization of the quantum link model Eq.~\eqref{eq:1dqlm} with $L=20$, from an initial state Eq.~\eqref{eq:inogauss2} with Gauss' law (bottom 
    strip) chosen such that some links remain frozen. This decouples two subsystems, which have no correlations or entanglement even in the long-time limit or in the Krylov-restricted thermal ensemble (top strip). Purple color denotes $\langle G_n \rangle = 3$.
    }
    \label{fig:QLMThermalizing2}
\end{figure}

The above example denotes a case in which the physical space is fully dynamically connected, and entanglement can be created between all pairs of sites. Some Gauss' law sectors can completely disconnect different subsystems. This results in no transport between these sections of the Hilbert space and no correlations or entanglement. We illustrate this in Fig. \ref{fig:QLMThermalizing2}, where we consider the same set of parameters in a larger volume ($L=20$). The state is initialized in a product state where
\begin{align}
    S^z_n&\ket{\psi}=\left(\delta_{n2}-\delta_{n3}+\delta_{n12}-\delta_{n13}\right)\ket{\psi} \nonumber\\
    c^\dagger_nc_n&\ket{\psi}=\left(\frac{1-(-1)^n}{2}\right)\ket{\psi}.
     \label{eq:inogauss2}
\end{align}
The dynamics breaks down into two different components, yet thermalization can still be achieved within the proper Krylov sector, now obtained with $\beta=0.0015$.

We note that the volumes depicted here are too small to conclude that the system is thermalizing, and weak ergodicity breaking could be present in this model. However, the observed non-translation invariance and decoupling of subsystems can be analytically predicted and therefore persist in the large-volume limit.

The same phenomena have been observed in models with Hilbert space fragmentation, which lack gauge symmetry \cite{Sala_2020, Moudgalya_2022}.

\end{document}